\pdfoutput=1
\documentclass[11pt]{article}
\usepackage{moriond}
\usepackage{amssymb}

\def\Journal#1#2#3#4{{#1} {\bf #2}, #3 (#4)}

\def\PLB{{\em Phys. Lett.}  B}
\def\PRL{\em Phys. Rev. Lett.}
\def\PRD{{\em Phys. Rev.} D}

\def\be{\begin{equation}}
\def\ee{\end{equation}}
\def\bea{\begin{eqnarray}}
\def\eea{\end{eqnarray}}

\newcommand{\Photo}{}

\begin{document}
\vspace*{4cm}
\title{Fundamental Physics with Imaging Atmospheric Cherenkov Telescopes}

\author{ A. Moralejo Olaizola }

\address{Institut de F\'{\i}sica d'Altes Energies, \\ 
Edifici Cn, Universitat Aut\`onoma de Barcelona, 08193 Bellaterra, Barcelona}

\maketitle\abstracts{Ground-based gamma-ray astronomy experienced a
  major boost with the advent of the present generation of Imaging
  Atmospheric Cherenkov Telescopes (IACTs) in the past decade. Photons
  of energies $\gtrsim 0.1$ TeV are a very useful tool in the study of
  several fundamental physics topics, which have become an important
  part of the research program of all major IACTs. A review of some
  recent results in the field is presented.
}

\section{Introduction}
Imaging Atmospheric Cherenkov Telescopes (IACTs) are currently the most
sensitive instruments for the observation of the Universe in the Very
High Energy band of the electromagnetic spectrum (VHE, $E_\gamma \sim$
0.1 - 100 TeV). There are now
three major IACT arrays in operation: H.E.S.S. in Namibia, MAGIC in
the Canary island of La Palma, and VERITAS in Arizona
(see\cite{holder} for a recent review of the field). Typical 
performance parameters of the current generation of 
IACTs are an energy threshold between $\simeq 20$ and 100 GeV, an
angular resolution O(0.1$^\circ$), an energy resolution of $\simeq
15\%$ (both of them energy-dependent), and an integral flux
sensitivity for point-like sources of about $1\%$ of the Crab Nebula
flux (or $\simeq 1.2 \times 10^{-12}~cm^{-2}s^{-1}$ above 300 GeV) in
25 hours of observation. The projected {\it Cherenkov Telescope Array} 
(CTA\cite{ctalamanna}) is expected to provide, by the end of this decade,
an order of magnitude improvement in sensitivity over existing
facilities. 
\par
Several topics related to fundamental physics can be addressed with
IACTs; here we review the results obtained in the past few years in
three of these areas: tests of the invariance of the speed of light,
the search for gamma rays from dark matter annihilation, and
the search for signatures of the existence of axion-like
particles. The fundamental physics prospects for CTA are presented
elsewhere in these proceedings\cite{moulin}.

\section{Testing the invariance of the speed of light}
Some quantum gravity (QG) theories predict violation of Lorentz
invariance (LIV) which, among other consequences, could result in
an energy dependence of the speed of
light\cite{amelino}. This effect would be suppressed by some large
QG energy scale, of the order of the Planck mass $m_{P}$ (or below, in
some models), so that the speed of light as a function of energy would 
behave as $v(E) = c ~ (1 \pm E/M_{QG1} \pm (E/M_{QG2})^2
\pm...)$. The observations with IACTs of  rapidly varying VHE emission from
active galactic nuclei (with flux-doubling timescales down to few
minutes) in a wide -O(TeV)- energy range, have not produced to date
any convincing evidence \footnote{MAGIC reported\cite{magic501} a small
  hint of energy-dependent time shift in the light curve of Mrk 501;
  it has to be noted that such delays, if confirmed, may also have
  less exotic explanations.} of this phenomenon\cite{magicliv,hessliv},
but have been used to set constraints on the values of $M_{QG1}$ and
$M_{QG2}$, for 
the cases of dominating linear and quadratic term respectively. The
best current limits, from H.E.S.S. observations of the blazar PKS
2155-304, are $M_{QG1} > 0.172 ~ m_{P}$  and $M_{QG2} > 5.2\times
10^{-9} ~ m_{P}$. The latter, though far from the Planck
scale, is the most constraining limit on the quadratic term obtained
by any technique, thanks to the long {\it lever arm} in energy
provided by IACTs. For the linear term, the Fermi-LAT limits from the
observation of Gamma-Ray Bursts (GRBs)\cite{fermiliv,vasileiou} are
the most constraining ones 
($M_{QG1} > 1.2 ~ m_{P}$). Since the relevant observable
is the {\it accumulated} photon delay upon arrival on Earth, very
distant sources are the best candidates for LIV tests; unfortunately,
the {\it horizon} for IACTs is limited by the non-perfect transparency
of the Universe to VHE radiation (see \S\ref{gammaprop}). This,
together with the limited field of view (FoV) of IACTs, makes the
detection of GRBs in the VHE range a challenging
task - yet to be accomplished. A promising alternative\cite{otte} for
LIV searches is the observation of 
pulsars, after the detection by VERITAS\cite{veritascrab} and
MAGIC\cite{magiccrab} of a power-law tail in the VHE emission of the
Crab pulsar. The smaller distance would be compensated by the fast
variability ($\sim$ms), which allows a more precise measurement of
possible delays, and by the possibility of accumulating long
observation times (limited in all other cases by the duration of the
flaring states).

\section{Indirect dark matter searches}

Weakly Interacting Massive Particles (WIMPs), with masses in the GeV -
TeV range are promising dark matter (DM) candidates, on the grounds of
the so-called WIMP miracle \cite{wimpmiracle}, i.e. the fact that, for a weak-scale
annihilation cross-section, their present relic abundance could
roughly account for the DM density inferred from cosmological
observations. WIMPs are part of various extensions of the standard
model (SM) of particle physics (e.g. supersymmetric models), and, upon 
annihilating into SM particles, are expected to produce gamma-rays
mostly via $\pi^0$-decay and final state radiation. The resulting gamma-ray
continuum spectrum would extend up to the DM particle mass, and hence
can reach the IACT energy band for $m_{DM}$ in the order of hundreds
of GeV or above. In this type of search, a drawback of IACTs with
respect to the lower-energy, space-borne telescopes, is their limited
FoV, which enforces the a-priori selection of few  
DM targets, and limits (through competition for observation time with
other programs) the exposure that can be accumulated on them. In the
case of annihilating DM, the most relevant parameter of a candidate
source is the volume integral, along the line of sight, of the squared
DM density over distance squared, often called {\it astrophysical
  factor}, which enters linearly in the computation of the expected
gamma-ray flux. Other desirable features are an angular extension well
below the telescope FoV, to facilitate evaluation of the background
due to cosmic-ray initiated air showers, and the lack of nearby
conventional astrophysical gamma-ray sources.

\subsection {Limits from dwarf spheroidal galaxies}
\begin{figure}
\begin{minipage}{0.50\linewidth}
\centerline{\includegraphics[width=0.98\linewidth]{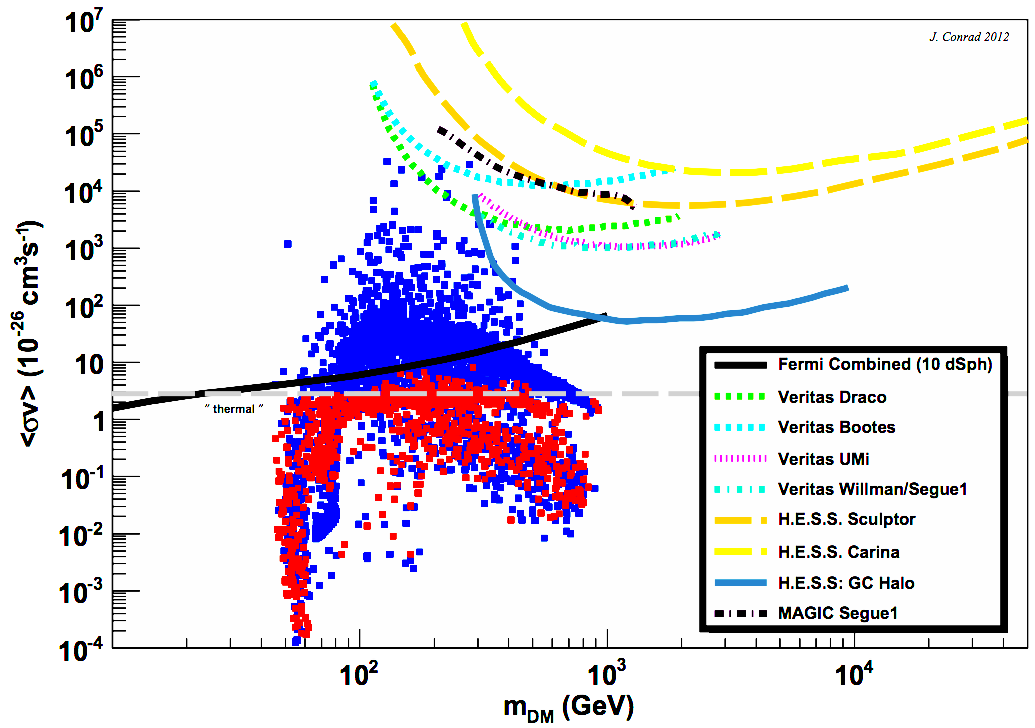}}
\end{minipage}
\hfill
\begin{minipage}{0.50\linewidth}
\centerline{\includegraphics[width=0.92\linewidth]{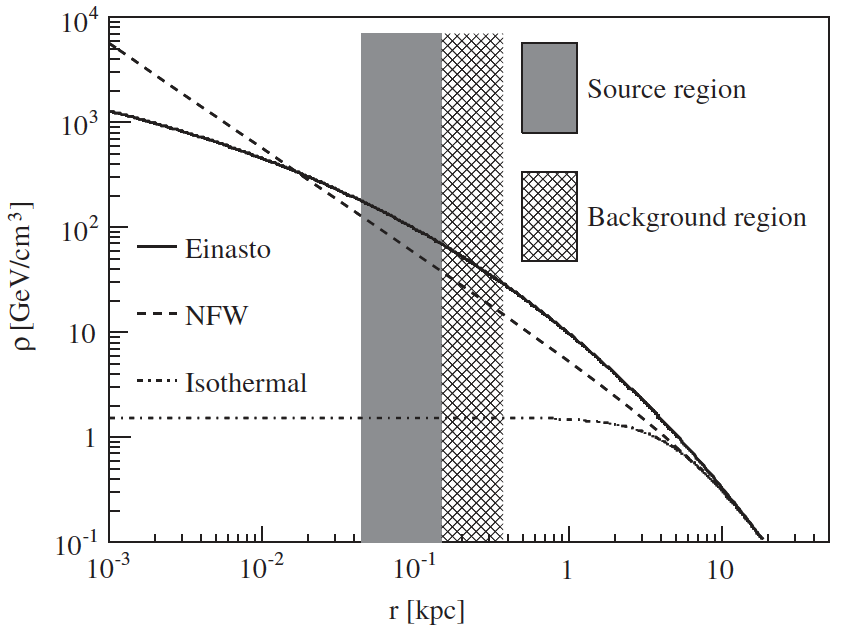}}
\end{minipage}
\vspace*{-0.3cm}
\caption{Left pad: compilation by Conrad \protect\cite{conrad} of
  various limits on $\langle \sigma_{ann} v \rangle$ by IACTs compared
  to the stacked analysis of 24 months of Fermi-LAT observations of 10
  dwarf spheroidal galaxies \protect\cite{fermidwarves}. Pure $b
  \bar{b}$ or $b \bar{b}$-dominated annihilation channel is
  assumed. The squares show models of phenomenological Minimal
  Supersymmetry, red ones being those assuming thermal DM production
  and with relic density consistent with WMAP
  measurements\protect\cite{oldfermidwarves}. Right pad: Milky Way DM
  profiles used in the H.E.S.S. search for an annihilation signal from
  the galactic halo\protect\cite{hesshalo}, and definition of the
  signal and background regions.}
\label{DMlimits}
\end{figure}

The above wish list has made of dwarf spheroidal galaxies orbiting our
own galaxy the most popular targets for indirect DM searches with
IACTs. All three major IACT arrays have conducted observational
campaigns on several of these objects, and no significant excess of
gamma rays has been observed from any of them (for the most recent 
results, see refs.\cite{magicsegue,hessdwarves,veritassegue}). Flux
upper limits 
can be transformed into an upper limit in the velocity-averaged
DM annihilation cross section, $\langle \sigma_{ann} v \rangle$,
for a given energy-dependent gamma-ray yield per annihilation
$dN_\gamma/dE(E)$. The latter is the result, in a specific particle
physics model (e.g. a given realization of supersymmetry), of the
sum of all possible annihilation channels. Alternatively, one can
assume 100$\%$ annihilation into a single channel, e.g. xx
$\rightarrow b \bar{b}$, and hence obtain
upper limits on  $\langle \sigma_{ann} v \rangle$ for that
specific channel. Even for the channels with highest gamma-ray yield
in the VHE range, the current IACT limits from dwarf spheroidal
observations are two to three orders of magnitude above the expected 
$\langle \sigma_{ann} v \rangle \lesssim 3 \times 10^{-26}~cm^3 s^{-1}$ for 
thermally produced DM (see left panel of fig. \ref{DMlimits}). It must
be noted that the estimated {\it astrophysical factors}, that are
needed to translate observations into DM constraints, have quite large
statistical and systematic uncertainties, which on one hand undermines
the robustness of the limits, but on the other may be used to argue
for the continuation of the observations, in the hope that detection
might be closer than expected. The same can be said of possible {\it
  boosts} of the gamma-ray flux induced by particle physics
effects, like additional contributions from internal {\it bremsstrahlung}
\cite{bergstrom} or the {\it Sommerfeld enhancement} \cite{sommerfeld}
of the cross section, in models where DM particles have long-range
interactions besides gravity.  

\subsection {Limits from observations of the galactic halo}

Flux-wise, the center of our own galaxy should be, for observers on
Earth, the brightest source of gamma rays from DM
annihilation. However, the presence of astrophysical gamma-ray
backgrounds (diffuse emission from cosmic ray interactions and a
strong source coincident with the position of the central black hole
Sgr A*) makes this a challenging region in the search for DM
annihilation. The situation improves as one gets away from the galactic
center, but then IACTs face the problems associated to the determination
of the background in the search for a faint diffuse gamma-ray excess
which spans (and hardly varies across) the whole FoV of the
telescopes. Instead of trying to set a limit on the absolute DM
annihilation flux in this central part of the galactic halo, the
H.E.S.S. collaboration \cite{hesshalo} looked for systematic
differences between the diffuse background rates (after masking all known
astrophysical sources) in two different ranges of galactocentric
distance (see fig. \ref{DMlimits}, right), with a careful selection of
the signal and background regions to ensure that they were completely
equivalent in terms of instrumental gamma-ray acceptance. In 112 hours
of live time, no significant excess was found in the signal
region. Under the assumption that the galactic DM halo follows a
Navarro-Frenk-White or an Einasto profile, the resulting limits
(fig. \ref{DMlimits}, right) on $\langle \sigma_{ann} v \rangle$ are
the best obtained by IACTs to date, and are just one order of
magnitude away from constraining the relevant part of the WIMP
parameter space. Note however that the 2-year Fermi-LAT observations
of dwarf spheroidals \cite{fermidwarves} provide the best constraints
up to DM masses as high as 1 TeV (thanks to its large FoV and duty cycle,
Fermi-LAT achieves much higher exposure than IACTs; besides, most of the
annihilation photons would be emitted at energies well below the DM
particle mass, within Fermi-LAT's range). The H.E.S.S. galactic halo
observations were also used recently \cite{hesslines} to set limits on
DM annihilation lines or other narrow spectral features in the energy
range 0.5 - 25 TeV.
\par
Clusters of galaxies have also been targeted by IACTs
\cite{magicperseus,veritascoma,hessfornax}, but currently 
provide weaker DM constraints than either dwarf spheroidals or the
galactic halo, with the additional complication of potential gamma-ray
{\it contamination} from active galaxies in the cluster and from
cosmic-ray interactions.

\section{Search for axion-like particles}

Axion-Like Particles (ALPs) are hypothetical spin-0 bosons with a
2-photon interaction vertex. They are a generalization of the axion
which would 
result from the spontaneous breaking of the Peccei-Quinn symmetry
postulated to solve the strong CP problem \cite{peccei}. ALPs can
convert into photons and vice-versa in the presence of an electric or
magnetic field (Primakoff effect), a process which could enable the
direct detection of ALPs in experiments like CAST \cite{cast} and
ADMX \cite{admx}. The existence of ALPs could also affect the
propagation of photons over cosmological distances. They were once
invoked as an alternative explanation for the dimming of type Ia
supernovae without resorting to cosmological acceleration \cite{snia}, 
as well as to account for the observation, by the AGASA
experiment, of an excess of cosmic rays of energies above the GZK
cutoff \cite{supergzk}. In the latter case, super-GZK events were
assumed to be 
ultra-high energy photons which convert to ALPs through interaction
with intergalactic magnetic fields, thus evade suppression via
e$^+$e$^-$ pair-production against the extragalactic radio background, 
and finally convert back to photons in the vicinity of the
Earth. However appealing, this solution was rendered unnecessary when
newer data 
from HiRes \cite{hires} and the Pierre Auger Observatory \cite{auger}
showed the presence of the 
expected GZK suppression in the cosmic ray spectrum. But the idea
that ALPs could play an important role in photon propagation in the
Universe would soon revive in the context of VHE astronomy. 

\subsection {VHE gamma-ray propagation \label{gammaprop}}
Propagation of VHE photons over intergalactic distances is hindered by the
presence of the so-called Extragalactic Background Light (EBL), an
ubiquitous radiation in the UV to IR wavelength range, which
results from the thermal emission by stars and dust in galaxies
throughout the history of the Universe. For center-of-momentum energies
above 2 $m_e$, VHE and EBL photons can interact to produce e$^+$e$^-$
pairs, a process which induces an energy-dependent depletion of the
VHE gamma-ray flux from distant sources. The flux suppression
increases with the gamma-ray energy, and sets a limit to the size of
the Universe observable in the VHE range, often referred to as the
``gamma-ray horizon''. 
\par
Direct measurements of the EBL are challenging due to the strong
foreground emission, mainly from zodiacal light. Robust lower
limits have been derived by integrating the contribution of resolved
galaxies in deep-field optical and infrared
observations \cite{madau,dole}. Upper limits to the EBL density were
derived from IACT observations (see e.g.\cite{meyerebl}), under the 
assumption that the intrinsic VHE spectra of BL Lac sources and other
blazars should have shapes allowed by the gamma-ray emission 
models, e.g. should not be much harder than $dF/dE \propto
E^{-1.5}$, and should become softer as energy increases. 
There were also claims \cite{da1,da2,masc} that some VHE spectra were
violating 
these constraints (or at least {\it in tension} with them), even when the
lowest possible EBL density was assumed in order to derive the
intrinsic spectra from the observations. Axion-Like Particles were
then proposed as a possible explanation of these anomalies (see
\S\ref{anomalies}).
\par
The first actual indirect measurements (not upper limits) of the EBL
density using gamma-ray observations were published independently by
the Fermi-LAT \cite{fermiebl} and H.E.S.S.\cite{hessebl} collaborations
in 2012. In both cases, a 
number of gamma-ray spectra were combined, using certain assumptions
on the intrinsic spectral shapes, to build a single likelihood which
was maximized to obtain the most likely scaling factor for the optical
depth $\tau(E,z)$, whose energy- and redshift dependence was taken
from the Franceschini '08 EBL model \cite{franceschini} - FR08 in the
following. These 
works concluded that, within uncertainties, the EBL density at UV -
near IR frequencies was compatible with that of FR08 and other similar
models, and less than $\simeq50\%$ above the lower limits from galaxy counts.

\subsection {Propagation anomalies and ALPs\label{anomalies}}
\begin{figure}[hb]
\centerline{\includegraphics[width=0.85\linewidth]{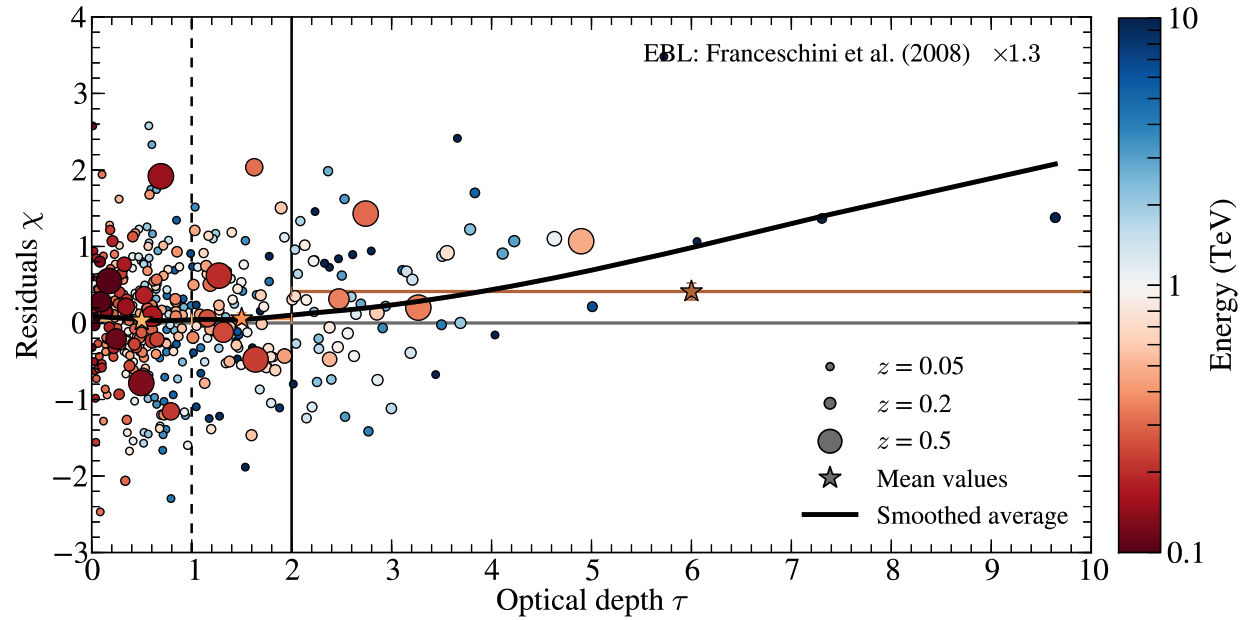}}
\vspace*{-0.3cm}
\caption{Relative residuals of measured VHE fluxes with respect to the
  expectations for reasonable intrinsic spectra and for the EBL
  density favoured by H.E.S.S. observations (taken from
  \protect\cite{meyer2}). The horizontal axis indicates the
  optical depth for $\gamma \gamma \rightarrow e^+ e^-$. Spectral
  points in the $\tau > 2$ regime lie in average above the
  expectations.\label{anomaly}
} 
\end{figure}

There have been several works \cite{da1,da2,masc} claiming that VHE
observations of some sources indicate that the Universe is more
transparent to gamma-rays than expected from ``low EBL'' models like
FR08 and others, in a sort of revival of the {\it TeV-IR
crisis} \cite{irtev_crisis} triggered by the 1997
observations of Markarian 501 by the HEGRA array of Cherenkov
telescopes \cite{mrk501_1,mrk501_2}. The most recent of these
claims, by Meyer {\it et al} \cite{meyer1,meyer2}, makes use of a sample
of 50 VHE gamma-ray AGN spectra from the current and previous generation
of IACTs, and studies how the spectral points in the optically-thick
regime (i.e. those affected by significant absorption in the EBL)
deviate from the fluxes expected under some reasonable assumptions on
the EBL density and on the intrinsic spectral shape of the
gamma-ray emission. The authors find that, using the best-fit EBL
density from H.E.S.S. \cite{hessebl} ($\tau \simeq 1.3 \times
\tau_{FR08}$), the spectral points at $\tau > 2$ are in average above
the expectation, i.e. they show a smaller flux suppression than
anticipated (see fig. \ref{anomaly}). This excess seems to be
correlated with the optical 
depth $\tau (E, z)$, and not with the energy of the spectral points,
hence suggesting the anomaly is a propagation effect, rather than
being related to the intrinsic source spectra. The statistical
significance of the anomaly is 3.5 standard deviations
\cite{meyer2}. They term this effect 
``pair production anomaly'', and speculate that it might be due to
conversion of gamma rays into (EBL-immune) axion-like particles and
{\it vice versa} in the magnetic fields traversed by the radiation.
\par
\begin{figure}
\centerline{\includegraphics[width=0.65\linewidth,height=7cm]{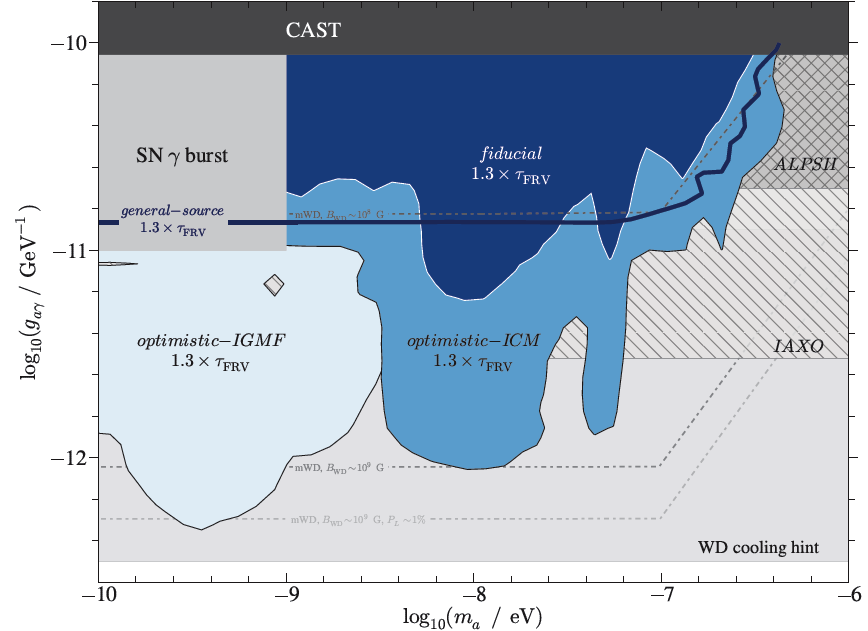}}
\vspace*{-0.3cm}
\caption{ALP parameter space (coupling constant
  vs. mass) from {\it Meyer et al \protect\cite{meyer3}},
  showing in different shades of 
  blue the regions in which ALPs might account for the so-called {\it
    pair production anomaly}, under different assumptions on the
  intervening magnetic fields.\label{ALPlowerlimits}
} 
\end{figure}

Under that assumption, they present in a separate
paper \cite{meyer3} {\it lower limits} to the photon - ALP coupling
constant $g_{a\gamma}$ as a function of the ALP mass. Since the
conversion of photons into ALPs depends on the magnetic fields in the
space between the source and the observer, several different scenarios
were considered for the source, its environment and the intergalactic
magnetic field. Conversions in the galactic magnetic field were also
included in the framework. For each of the B-field scenarios and
scanned ALP masses, the minimum value of $g_{a\gamma}$ to reproduce
the observed anomaly was computed
(fig. \ref{ALPlowerlimits}). Although this is presented as a {\it
  lower limit} to $g_{a\gamma}$, this is certainly not a ``limit'' in the same
sense as the upper limits from direct axion searches. The existence of
ALPs with lower coupling constants, or even the non-existence of ALPs,
is not {\it forbidden} by these observations. What
fig. \ref{ALPlowerlimits} really shows is the region of the parameter
space in which ALPs would be a viable explanation for the pair
production anomaly.
\par
It must be remarked that the statistical significance of the {\it
  pair production anomaly} is just 3.5 $\sigma$, and besides, there
are a 
number of possible systematic effects which may be contributing to
it. For example, in steep spectra, there is a significant spill-over
of events towards larger energies, given the limited energy resolution
of the IACT technique ($\Delta E / E \simeq 15\%$). The correction of
this effect requires a good matching of the data and the Monte Carlo
(MC) simulations used in the calculation of the instrument response
(something difficult to achieve, given the crucial role of the
atmosphere in the IACT technique). Since energy reconstruction is
trained on MC, any mismatch will likely result in worse energy
resolution in the real data as compared to the simulations, hence in
larger event spill-over in the data, which will not be fully corrected
by the simulation-based response function. Another problem comes from
the fact that the highest-energy points of VHE spectra are naturally
biased towards higher fluxes: for an average flux slightly below the
instrument sensitivity, positive fluctuations, of the signal or of
the background, will make the point to become part of the measured
spectrum, and the estimated flux will then have a positive bias. The
corresponding negative fluctuations, on the contrary, would not be
present in the spectra. The authors of the first paper \cite{meyer1}
discuss these sources of systematics and conclude that in the worst
case they would reduce the significance of the anomaly from 4.2
$\sigma$ to 2.6 $\sigma$ \footnote{In the 2012 
  paper by Horns and Meyer \cite{meyer1} the so-called ``minimal EBL
  model'' was used, and the significance of the anomaly was 4.2
  $\sigma$. For the updated 3.5 $\sigma$ result \cite{meyer2}, using
  the scaled FR08 EBL, the effect of systematics is not reported, but
  assuming it to be similar, it would bring the significance down to
  around 2 $\sigma$.}.

\par
\begin{figure}
\centerline{\includegraphics[width=0.8\linewidth]{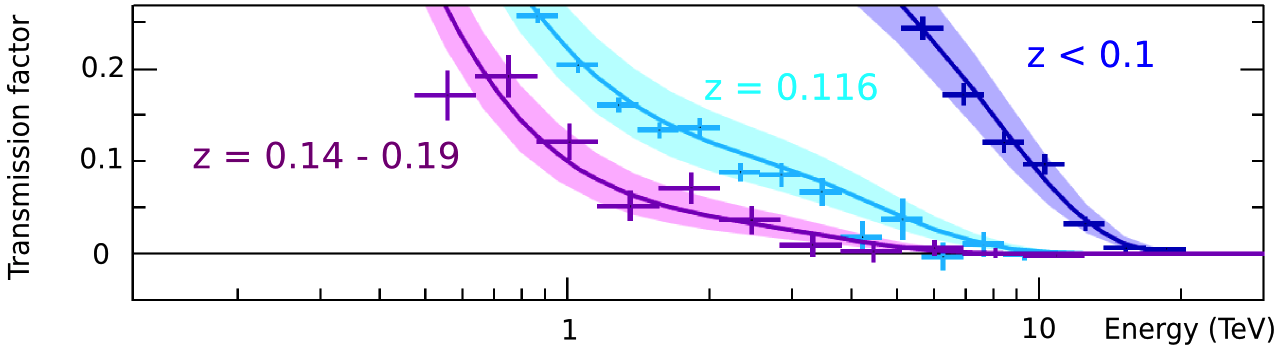}}
\vspace*{-0.3cm}
\caption{Transmission factor vs. gamma-ray energy for the
  best joint likelihood fit of EBL density and intrinsic
  spectra of a sample of very high quality H.E.S.S. observations of
  bright blazars. Adapted from \protect\cite{hessebl}. Three ranges of
  redshift are shown separately. Even for optical
  depth $\tau > 3$ (transmission $< 0.05$) the data (points with
  error bars) do not seem to be systematically above the expectations
  (solid lines and shaded regions).\label{HESSebl}}   
\end{figure}

On the other hand, in the 17 high-quality VHE spectra from 7 sources
used by the H.E.S.S. collaboration in the measurement of the EBL
density, there seems to be no hint of anomalies for any of the
spectral points, even at large optical depths (see
fig. \ref{HESSebl}). It might be argued, indeed, that the high-$\tau$
points enter the {\it fit} which determines the EBL density, but the
energy- and redshift dependence of $\tau$ are fixed to those 
of the FR08 model, so anomalous high-$\tau$ points could not possibly
pull the fit without worsening the agreement at lower $\tau$. As
mentioned above, the best-fit normalization, within uncertainties, is
perfectly compatible with the FR08 value. Since the hypothesized
ALP-gamma mixing would depend on the magnetic field structure between
source and observer \cite{mirizzi}, it is just possible that for those
particular sources, by chance, the net effect of the ALPs is
negligible. A plausible alternative to reconcile both results without
ALPs is to blame the anomaly fully on experimental systematics which
may be absent in high-quality spectra like those used in the
H.E.S.S. EBL measurement. 
\par
Another recent work in which ALPs are proposed as a solution for an
observed anomaly in VHE data is the paper \cite{tavecchio} by
Tavecchio {\it et al}, which addresses the difficulties in modelling
the observations by MAGIC \cite{pks1222} of the quasar PKS
1222+216. The fast variability of this object indicates a compact
emission region close to the central engine, i.e. in a very
$\gamma$-ray opaque environment, due to photon-photon interaction in
the dense UV fields originated in the broad line region. While ALPs
are, once again, a possible way to reduce the optical depth and hence
explain the observations, alternative models exist (see discussion in
\cite{tavecchio}) which do not require new physics.
\subsection{Spectral irregularities as a signature of photon - ALP mixing}
It has been noted \cite{mirizzi,burrage,wouters1} that, due to the
turbulent nature of the intergalactic magnetic fields (IGMF), the
effects of 
ALPs on gamma-ray propagation, and in particular, the possible
reduction of the EBL-induced flux suppression, will depend on the
detailed magnetic field structure along the beam path, and will be
impossible to predict for a single source. Even under the assumption
of a certain IGMF intensity (or spectrum of intensities) and
coherence length, only the average {\it effect} on a large number of
sources can be predicted for a given ALP scenario. Wouters {\it et al}
\cite{wouters1} propose an alternative method which can be applied to
individual VHE spectra, namely to look for spectral {\it
  irregularities} resulting for the strong energy dependence of the
ALP$\rightarrow \gamma$ conversion probability in the so-called {\it weak
mixing regime}, at energies close to the threshold of the
process. This method has already been applied by the
H.E.S.S. collaboration, and results are presented elsewhere in these
proceedings \cite{wouters2}.

\section{Conclusions}

With the current generation of IACTs, astronomy in the VHE band has
reached its maturity, and is providing a wealth of data which allow to
address, besides the traditional topics of high-energy astrophysics, a
number of questions in the field of Fundamental Physics. Despite
some interesting hints, no evidence for new phenomena has been found
to date. Nonetheless, IACTs are already providing competitive
constraints which foster the hopes set in the next-generation
ground-based gamma-ray telescope, CTA\cite{ctafunda}.

\section*{Acknowledgments}

I would like to thank the organizers for the invitation to participate
in this very interesting conference, and for the excellent
organization. I also want to acknowledge the support of the MultiDark 
CSD2009-00064 project of the Spanish Consolider-Ingenio 2010 programme
of the Spanish MINECO.

\end{document}